\documentclass[sigconf]{acmart}
\AtBeginDocument{%
  }

\copyrightyear{2025}
\acmYear{2025}
\setcopyright{cc}
\setcctype{by}
\acmConference[ICHEC 2025]{Proceedings of the 2025 International Conference on Human-Engaged Computing}{November 21--23, 2025}{Singapore, Singapore}
\acmBooktitle{Proceedings of the 2025 International Conference on Human-Engaged Computing (ICHEC 2025), November 21--23, 2025, Singapore, Singapore}
\acmPrice{}
\acmDOI{10.1145/3786995.3787011}
\acmISBN{979-8-4007-2234-9/2025/11}

\usepackage{tabularx}
\usepackage{booktabs}

\usepackage{tikz}
\usetikzlibrary{arrows.meta,positioning,calc,fit,backgrounds,shapes.geometric}
\usepackage{pgfplots}
\pgfplotsset{compat=1.18}

\usetikzlibrary{shapes.misc, arrows.meta, shadows}

\usepackage{graphicx}

\begin{document}

\title[Temporal Drift in Privacy Recall]{Temporal Drift in Privacy Recall: From Verbatim Loss to Gist-Based Overexposure}


\author{Haoze Guo}
\email{hguo246@wisc.edu}
\orcid{0009-0009-5987-1832}
\affiliation{%
  \institution{University of Wisconsin - Madison}
  \city{Madison}
  \state{Wisconsin}
  \country{USA}
}

\author{Ziqi Wei}
\email{zwei232@wisc.edu}
\orcid{0009-0009-7541-8376}
\affiliation{%
  \institution{University of Wisconsin - Madison}
  \city{Madison}
  \state{Wisconsin}
  \country{USA}
}

\renewcommand{\shortauthors}{Guo et al.}

\begin{abstract}
With social media content traversing the different platforms, occasionally resurfacing after periods of time, users are increasingly prone to unintended disclosure resulting from a misremembered acceptance of privacy. Context collapse and interface cues are two factors considered by prior researchers, yet we know less about how \emph{time-lapse} basically alters recall of past audiences destined for exposure. Likewise, the design space for mitigating this temporal exposure risk remains underexplored. Our work theorizes \emph{temporal drift in privacy recall} as verbatim memory of prior settings blowing apart and eventually settling with gist-based heuristics, which more often than not select an audience larger than the original one. Grounded in memory research, \emph{contextual integrity}, and usable privacy, we examine why such a drift occurs, why it tends to bias toward broader sharing, and how it compounds upon repeat exposure. Following that, we suggest provenance-forward interface schemes and a risk-based evaluation framework that mutates recall into recognition. The merit of our work lies in establishing a \emph{temporal awareness of privacy design} as an essential safety rail against inadvertent overexposure.
\end{abstract}




\maketitle

\section{Introduction}
\begin{figure}[!t]
  \centering
  \includegraphics[width=\linewidth]{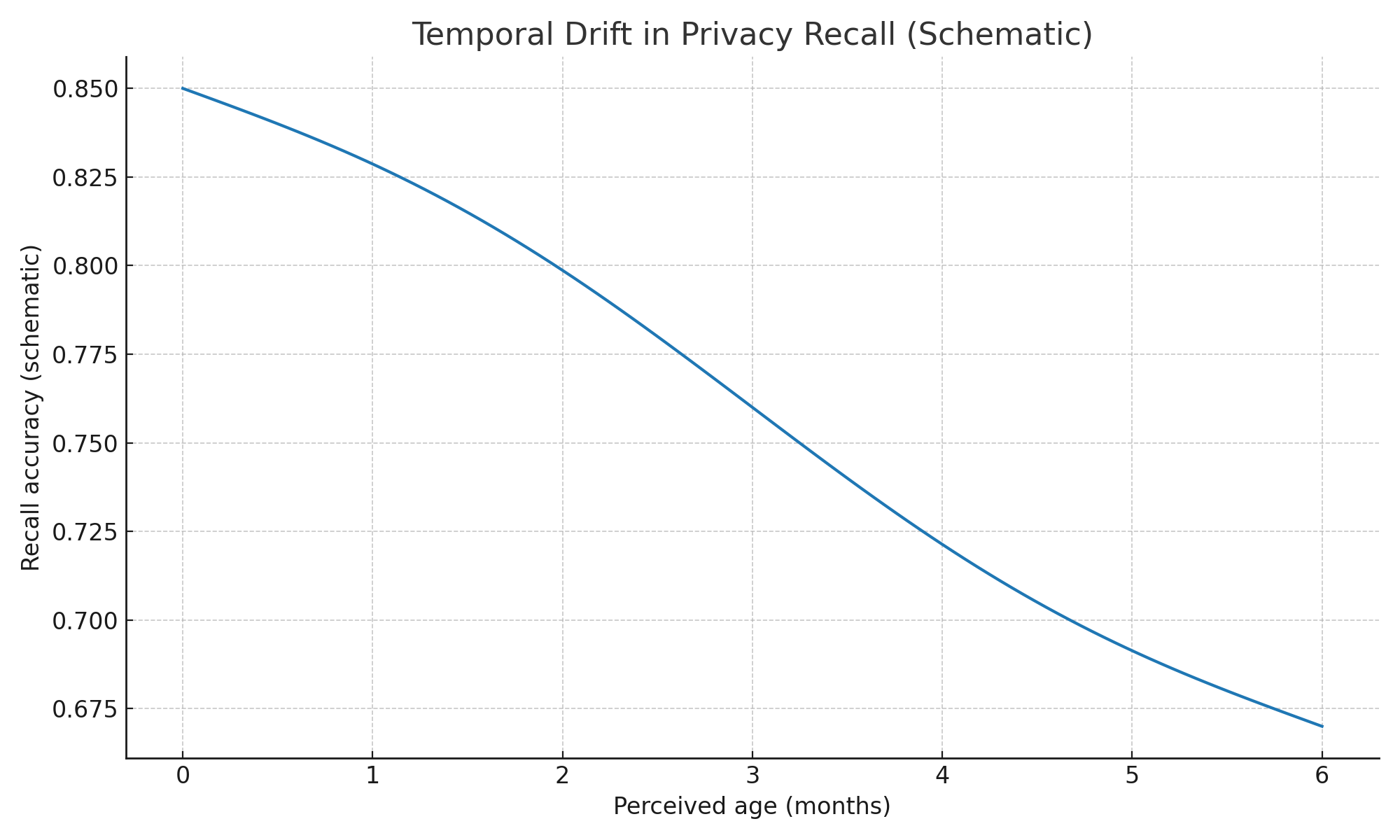}
  \caption{Temporal drift schematic: recall specificity declines as perceived age increases, raising the chance of scope misremembering.}
  \Description{A simple line chart showing a gentle decline in recall accuracy as perceived age increases.}
  \label{fig:drift}
\end{figure}

\subsection{Motivating Vignette}
Six months after posting end-of-semester photos to a ``Friends'' audience, T. drags one image into a new composer. A bold timestamp reads ``6 months ago.'' Asked ``Who could see this then?,'' T. answers ``Friends-of-Friends'' with high confidence—wrong by one step. That subtle inflation, repeated across reuse flows, creates \emph{overexposure risk} for sensitive content.

\subsection{Problem and Stakes}
Failures in security and privacy are often beyond access control and instead have their roots in \emph{human memory} about the controls. With longer content lifetimes and increased cross-platform reuse, memory-related errors pile up against risk. We use the concept of \emph{temporal drift in privacy recall} to name a systematic tendency wherein the remembered scope-audience changes with elapsed time. Drift, then, affects \emph{accuracy} (are we right?) and \emph{direction} (if wrong, are we more open or more closed than the truth?). We discuss background and formalize drift; then we glide into light-touch, human-centered mitigations for drift that supplement the technical access controls.

People reshare or cross-post older content all the time. The audience selected at creation might be forgotten months later as context changes. This gap through\emph{temporal drift in privacy recall} opens up unintended disclosure risks, complementing classic \emph{context collapse} \cite{boyd2011tweet,Vitak2012ContextCollapse}. We focus on a ubiquitous visual cue: \emph{perceived age}-and draw from it design implications for making past audiences recognizable at the moment of reuse.

\subsection{Operational Definition and Threat Model}
We use the term \emph{temporal drift in privacy recall} to refer to a consistent difference between the audience for a post the poster created and the audience they recall at reuse time as time passes. Formally, let $A_0$ be the original audience and $\hat{A}_t$ the remembered audience when the item is resurfaced at perceived age $t$. A drift event occurs when $\hat{A}_t \neq A_0$, with \emph{directionality} measured along a simple lattice \emph{Private}$\!\to$\emph{Friends}$\!\to$\emph{Public} that approximates common scopes across platforms. This construction is not a claim about full audience ontologies, but a useful pragmatic scaffold for thinking through one-step overexposure errors when verbatim details decay faster than gist and when source does not match \cite{Ebbinghaus1913Memory,Johnson1993SourceMonitoring,Schacter1996Illusions}. In contextual-integrity terms, the parameter “who could see this then” becomes latent unless interfaces restore it \cite{Nissenbaum2004ContextualIntegrity}. Our threat model treats an overexposure as any case where $\hat{A}_t$ is strictly broader than $A_0$ (e.g., \emph{Friends} $\to$ \emph{Public}), with risk amplified by context collapse and misleading salience cues (large age labels, absent provenance) \cite{boyd2011tweet,Vitak2012ContextCollapse,Brandtzaeg2018TimeCollapse}. The foregoing helps to motivate a design differential between recall and recognition and a cost-sensitive policy for when and if to intervene when poster is reused.

\subsection{Scope and Assumptions}
Beyond isolated mishaps, drift compounds: resurfacing features encourage annual ``memories,'' cross-app editors strip provenance, and collaborative tools remix items into new contexts. Without explicit audience provenance, people lean on typicality (\emph{``how I usually share''}) rather than specific memory traces. In systems terms, the state variable ``original audience'' is not made observable at reuse time, so users infer it, often upward along the \emph{Private}$\!\to$\emph{Friends}$\!\to$\emph{Public} order. We use this three-level scaffold as a tractable lens for directionality.

In sum this paper make the following contributions: (1) a mechanism account linking perceived age, memory dynamics, and audience miscalibration; (2) actionable design patterns that keep audience provenance salient during reuse with minimal friction; and (3) a decision-analytic risk model and non–user-study evaluation blueprint.

And the study then bring into four primary research questions are:

\begin{itemize}
\item \textbf{RQ1:} How can interfaces convert privacy \emph{recall} into \emph{recognition} during reuse with minimal friction?
\item \textbf{RQ2:} How can teams evaluate such patterns \emph{without} new user studies, using log-only, privacy-preserving methods?
\item \textbf{RQ3:} Why does perceived age reduce accurate recall of past audiences?
\item \textbf{RQ4:} When recall is wrong, why are errors directionally biased toward broader exposure?
\end{itemize}

\subsubsection{Claim scope.}
In this paper we treat recall errors tend to bias the other way, biased toward topical over-exposure. This bias is theoretically supported by source-monitoring and gist-over-verbatim poetry aspects, but we are not claiming this is an unbreakable law. In practice, teams using the bias should treat it as a \emph{testable prior} rather than a fact: when priors are uncertain, the optimal policy is to set bounds conservatively and retain sensitivity bounds.

\section{Related Work}

In the literature on social media, the audience reasoning is entwined with the idea of an \emph{imagined audience}: a poster forms models of the sort of people who might see their content and then behave accordingly-if those mental models are often miscalibrated \cite{boyd2011tweet,Vitak2012ContextCollapse,Marwick2011ImaginedAudience,Litt2016ImaginedAudience,Bernstein2013InvisibleAudience}.  Due to \emph{context collapse}, heterogeneous groups of people-family, friends, work colleagues, etc.-are compressed into single broadcast venues, which makes it difficult to maintain types of scopes that must be occasionally nuanced. Even if technological aids are in play, e.g., friends lists, groups, restricted audiences, they demand one to keep up their attention just in time, often at the cost of gradual degradation due to changes in membership or defaults. It is, thus, a social as well as interface problem: the UI lets one choose to emphasize certain dimensions (such as elapsed time) and glaringly omit prior audience, leading to the formation of misinformed choices and decisions \cite{Rader2014Tracking,Eslami2016FolkTheories}.

Contextual integrity frames privacy as appropriate information flow—what attributes, transmitted under which principles, to which roles, in which contexts \cite{Nissenbaum2004ContextualIntegrity}. Audience provenance—``who could see this then''—is a core parameter of that flow. When resurfacing elides provenance, users must infer appropriateness without a key dimension. That omission is consequential when content migrates across apps with incompatible audience ontologies or when ``memories'' features highlight time but not the original recipients \cite{Gilbert2023Reuse,Brandtzaeg2018TimeCollapse}. Empirically, small changes in defaults and salience can shift disclosure \cite{Wang2014PrivacyNudgesFacebook,Stutzman2010FriendsOnly,Madejski2012PrivacySettings}; in reverse, omission of salient parameters can yield miscalibrated behavior.

\begin{table*}[!t]
\small
\centering
\begin{tabular}{p{2.8cm} p{6.2cm} p{5.2cm}}
\toprule
\textbf{Strand} & \textbf{Key idea} & \textbf{Design implication}\\
\midrule
Imagined \& invisible audiences & People misestimate who sees posts; mixed groups collapse into one venue \cite{boyd2011tweet,Litt2016ImaginedAudience,Bernstein2013InvisibleAudience}. & Make the actual prior audience visible at reuse time to correct mental models.\\[2pt]
Contextual integrity & Privacy = appropriate flows given roles \& transmission principles \cite{Nissenbaum2004ContextualIntegrity}. & Surface “who could see this then” as first-class provenance during reuse.\\[2pt]
Memory dynamics & Verbatim fades faster than gist; source misattribution \cite{Ebbinghaus1913Memory,Johnson1993SourceMonitoring,Schacter1996Illusions}. & Convert recall to recognition; show prior scope, not just age.\\[2pt]
Choice architecture & Small cues \& defaults steer disclosure \cite{Wang2014PrivacyNudgesFacebook,Johnson2003DefaultsSaveLives,Thaler2008Nudge}. & Use narrow-first defaults when provenance is missing; avoid dark patterns.\\
\bottomrule
\end{tabular}
\caption{Four strands motivating provenance-forward reuse: make the original audience legible to reduce temporal drift.}
\label{tab:literature-compact}
\end{table*}

Memory research makes the temporal dimension sharper. Verbatim details decay rapidly compared to gist \cite{Ebbinghaus1913Memory,Schacter1996Illusions}, and people often commit \emph{source-monitoring} errors—remembering that they shared but misattributing audience or channel \cite{Johnson1993SourceMonitoring,Brainerd2002FuzzyTrace,Roediger1995DRM,Schacter2007RememberingFuture}. When a large ``6 months ago'' label is visible and audience provenance is not, retrieval relies on habits (``I usually post publicly'') or norms (``family photos are for friends''), leading to one-step drift along typical audience orders. Photography and platform automation can further distort memory; for instance, offloading memory to photos or feeds changes what details people recall \cite{Henkel2014Photo,Sparrow2011Google}. Overconfidence compounds the problem: people can feel certain while being wrong \cite{Koriat2012SelfConsistencyConfidence,Moore2008Overconfidence}.

There need to be three ingredients for temporal drift: the salience of age, invisibility of provenance, and an audience lattice that allows for "one step more open" errors. Further, some remedies are suggested: Display the relevant parameters at the point of decision \cite{Egelman2010TimingPrivacyIndicators}; introduce ethical defaults \cite{Johnson2003DefaultsSaveLives,Thaler2008Nudge}; eschew dark patterns \cite{Gray2018DarkPatternsSide,Mathur2019DarkPatterns}; and help people form good folk theories by using outcome-based copy, not technical jargon \cite{Rader2014Tracking}. We take this into concrete flora of patterns and a risk-aware policy for when to show them.

\section{Discussion and Design Implications}
\subsection{Mechanism \& Core Patterns}
From the standpoint of temporal drift, it arises as a boundary-case within the recognition-versus-recall framework. If we think about the resurfacing of an old item as a \emph{recall} event, then it is..." For years, human-computer interaction wisdom has advised favoring recognition over recall because "recognition means less brain power and fewer chances to go wrong" \cite{Nielsen1994Usability,Sweller1988CognitiveLoad}. That seemingly innocent timestamp devoid of provenance violates that principle just when it should have been upheld \cite{Nissenbaum2004ContextualIntegrity,Rader2014Tracking,Eslami2016FolkTheories}. The temporary fix is simple: pass the audience parameter so that the user interface offers recognition with possible deferral \cite{Moreau2013PROVDM,Egelman2010TimingPrivacyIndicators}. It endorses autonomy as an expression from sharing while making sure there is no unwanted overexposure.

A first pattern is a portable \emph{provenance badge}: a short, readable label that travels with the media wherever it appears in reuse flows, for example, “Original audience: Friends — May 2024.” The badge answers the precise, situated question posters are likely to have (“who could see this then?”) and can be implemented as metadata read at render time \cite{Nissenbaum2004ContextualIntegrity,Bernstein2013InvisibleAudience,boyd2011tweet,Moreau2013PROVDM}. Tapping reveals a small, accessible sheet that preselects “Carry over Friends” and provides one-tap override to broaden or narrow \cite{Johnson2003DefaultsSaveLives,Thaler2008Nudge}. The interaction is designed for recognition, not search; it keeps routine reuse fast while making risk-relevant choices reversible and explicit \cite{Nielsen1994Usability}.

A second pattern involves the \emph{context-restoration prompt}, which surfaces only when it is needed most. Many platforms, ranging from \emph{WeChat} to \emph{Instagram}, can locally infer indecision; for instance, by detecting on-device hesitation or backtracking through the interface, but without logging raw behavior \cite{Tsai2011PrivacySealsISR,Leon2012OBANotices}. When an item is marked as \emph{Old}, and uncertainty is detected - or if the content category is sensitive - the UI jumps up briefly to confirm its prior audience with a narrow-first alternative \cite{Brandtzaeg2018TimeCollapse}. The copy never scolds or talks above the user in jargon but instead describes outcomes ("only your friends will see this") \cite{Rader2014Tracking,Eslami2016FolkTheories}. When user certainty appears to be very high, the UI stays silent unless the requested scope conflicts with provenance, in which case a gentle clarification pops up, with an easy option to preserve the original scope \cite{Koriat2012SelfConsistencyConfidence,Moore2008Overconfidence}.

The third pattern comes in for \emph{imports without provenance}. Cross-app reuse frequently strips or reinterprets audience information \cite{Gilbert2023Reuse}. In these situations, it is more ethical to start with the conservative scope while providing an easy way to broaden it than to widen the scope silently. Defaults do possess power; hence, they should be used to defend a user's rights, not to prey on them \cite{Johnson2003DefaultsSaveLives,Thaler2008Nudge}. When audience mappings turn out to be lossy or ambiguous, the interface should make the ambiguity explicit and let the poster decide \cite{Gray2018DarkPatternsSide,Mathur2019DarkPatterns}.

\subsection{Accessible Deployment, Portability, \& Governance}

Those patterns have to be accessible and international in implementation. Screen-reader labels should expose the badge text (e.g., `Original audience: Friends`) and the purpose for controls (e.g., `carry over original audience`). Labels, icons, and focus order must allow keyboard navigation; color contrast must consider common impairments; date formats must be localized; and right-to-left layouts need validation. Copies are outcome-oriented, not mechanism-oriented. They say what will happen (`only your friends will see this`) and not how the system judges it. Because users build folk theories about feeds and privacy controls \cite{Eslami2016FolkTheories,Rader2014Tracking}, correct and easy-to-understand explanations reduce the confusion and increase the users' trust.

The patterns are also portable beyond social media. Usually, documents are duplicated, compiled, or moved across spaces in a shared drive or an enterprise collaboration tool, where the default visibility settings differ. Showing labels such as "Original sharing scope: Your team — Feb 2024" and allowing one-click carry-over ease coordination failures and reduce interpersonal friction \cite{Lampinen2011Together,Petronio2002BoundariesOfPrivacy}. Narrow-first defaults and clear boundary markers protect against overexposure, particularly when materials persist for long periods, in LMSs or patient portals, where overexposure costs can be much greater.

Privacy labels are in order. Uncertainty detection should transpire and be discarded after gating strictly on the device; maintain only aggregate counters pertaining to evaluating interventions with strict opt-in consent and minimization \cite{Tsai2011PrivacySealsISR,Leon2012OBANotices}. Pattern libraries should articulate the rationale where escalations are concerned and should not use deceptive pre-selection or any architecture that nudges the user toward accepting a broader scope of use without their informed consent \cite{Gray2018DarkPatternsSide,Mathur2019DarkPatterns}. Foreground provenance where it is known; otherwise, plainly state the lack of provenance and default conservatively. If the provenance is sensitive (for instance, a prior share to a stigmatized group), its membership should not be revealed in an interface absent explicit user choice; instead, the badge can abstract to a neutral label such as "Limited audience—May 2024" with an affordance to reveal details.

\subsection{Edge cases}
Not all content or communities experience temporal drifting in a similar manner. Some items have unusually high downstream sensitivity (e.g., health-related postings, posted identities with stigma attached, content that was created and shared under distinct norms) such that even a minor inflation of scope can yield very high social costs. For these items, provenance badges should be designed as unequipped with the explicit labels of those items unless a user opts to broaden them, starting with something neutral (i.e., "Limited audience - May 2024") and then allowing the user to remove that constraint if they so wish. Defaults for restoration prompts should similarly follow the most narrow viable defaults, with an explicit reverse path toward broadening \cite{Petronio2002BoundariesOfPrivacy,Johnson2003DefaultsSaveLives,Thaler2008Nudge}. Another potential failure mode would happen when a user is predictably \emph{incorrect and overconfident}; when the user is overconfident, it can overshadow misrecall, simple users should simply nudge backwards along their last recalled norm (like, "This was originally shared with Friends, keep that as is, or broaden it?") , and once again avoid silently compliant relationships with users \cite{Koriat2012SelfConsistencyConfidence,Moore2008Overconfidence}. Finally, designers should take adversarial misuse into account; e.g, if a malicious actor were to strip provenance and re-upload the content causing a default to go public. Defensive heuristics included Narrow-first, for any acceptable post that does not have verifiable provenance within the universe, highlighting user uncertainty towards the poster and supporting single-touch reporting options across suspicious resharing flows \cite{Madejski2012PrivacySettings,Gray2018DarkPatternsSide,Mathur2019DarkPatterns}.

\section{Risk Model and Evaluation for Temporally Aware Privacy UX}

We formalize gating as a cost-sensitive decision over features $\mathbf{x}$ that yield estimates of $p_c$ and $p_o$, combined into a risk score $(1-p_c)p_o$, and compare expected benefit $c_o \Delta r$ to friction $c_f$; see
Fig.~\ref{fig:risk_pipeline}.

\begin{figure*}[!t] 
\centering 
\begin{tikzpicture}[font=\small, node distance=8mm and 8mm] 
\tikzset{ block/.style={draw, rounded corners=3pt, minimum width=22mm, align=center, inner sep=3pt}, decision/.style={diamond, draw, aspect=2, inner sep=1.5pt, align=center}, } 
\node[block] (x) {Features $\mathbf{x}$\\(age, sensitivity,\\uncertainty)}; \node[block, right=1.8cm of x] (pc) {Estimate $p_c(\mathbf{x})$\\(correct recall)}; 
\node[block, below=1.0cm of pc] (po) {Estimate $p_o(\mathbf{x})$\\(overexposure if wrong)}; 
\node[block, right=1.8cm of pc] (risk) {Risk score\\ $(1-p_c)p_o$}; \node[decision, right=1.8cm of risk] (th) {$c_o\Delta r > c_f$?}; \node[block, above right=6mm and 10mm of th] (prompt) {Show badge/sheet}; \node[block, below right=6mm and 10mm of th] (noprompt) {No prompt}; 

\draw[-{Triangle[length=2mm]}] (x) -- (pc); 
\draw[-{Triangle[length=2mm]}] (x) |- (po); 
\draw[-{Triangle[length=2mm]}] (pc) -- (risk); 
\draw[-{Triangle[length=2mm]}] (po.east) -| (risk.south); 
\draw[-{Triangle[length=2mm]}] (risk) -- (th); 
\draw[-{Triangle[length=2mm]}] (th) -- node[above]{Yes} (prompt); 
\draw[-{Triangle[length=2mm]}] (th) -- node[below]{No} (noprompt); 

\end{tikzpicture} 
\caption{Cost-sensitive gating: estimate $p_c$ and $p_o$ to form $(1-p_c)p_o$, then intervene when the risk-weighted benefit $c_o\Delta r$ exceeds friction $c_f$.}
\label{fig:risk_pipeline} 
\end{figure*}

\subsubsection{Decision-analytic model.} Designers and policy teams must decide \emph{when} to show which control and \emph{how} to justify the trade-offs between friction and risk reduction. A small decision-analytic model helps formalize those choices without requiring a lab study. Let $p_c(\mathbf{x})$ denote the probability that a poster recalls the prior audience correctly for a reuse event with features $\mathbf{x}$ (e.g., perceived age, topical sensitivity, on-device hesitation, device context). Let $p_o(\mathbf{x})$ denote the probability that, conditional on an incorrect recall, the error widens scope (overexposure). The expected \emph{per-event} overexposure risk is
\[
\mathbb{E}[R \mid \mathbf{x}] \;=\; \big(1 - p_c(\mathbf{x})\big) \cdot p_o(\mathbf{x}).
\]
A lightweight intervention (badge and/or confirmation sheet) changes these terms to $\tilde{p}_c,\tilde{p}_o$ by turning recall into recognition and preselecting a scope-preserving choice. The absolute reduction in per-event risk is
\[
\Delta r(\mathbf{x}) \;=\; \big(1 - p_c(\mathbf{x})\big)p_o(\mathbf{x}) - \big(1 - \tilde{p}_c(\mathbf{x})\big)\tilde{p}_o(\mathbf{x}).
\]
Intervening imposes a small friction cost $c_f$ (time, attention, abandonment probability) and averts a harm with expected cost $c_o$ (user harm, complaints, takedowns, moderation load). Show the control when
\[
c_o \cdot \Delta r(\mathbf{x}) \;>\; c_f,
\]
i.e., when risk-weighted benefit exceeds cost. This yields an interpretable, cost-sensitive threshold on the \emph{risk score} $\big(1 - p_c(\mathbf{x})\big)p_o(\mathbf{x})$. If the platform estimates or brackets $p_c,p_o$ from historical logs or literature-informed priors, the policy reduces to “intervene on \emph{Old}$\wedge$uncertain (or \emph{Old}$\wedge$sensitive) events first,” because those lie in the right tail of risk. When $c_o \gg c_f$, the optimal operating point favors sensitivity over specificity; when $c_o \approx c_f$, the policy can be more conservative \cite{DeGroot2004Optimal,Fawcett2006ROC}.

\subsubsection{Estimating $p_c(\mathbf{x})$ and $p_o(\mathbf{x})$.} Without that ground-truth label of "correct recall," the model sells for useful purposes. The teams are able to infer proxies: if there existed previously an audience distinct from the outgoing one (an event of scope expansion), if there is a high amount of hesitation on \emph{Old}-stuff, or complaints coming downstream in the form of tickets about unintended exposure. If instrumented well and privacy-by-design is baked in, the team can explore several operating points by adjusting an uncertainty threshold and watching aggregate-level guardrails such as completion rate, time-to-post, or override share. It is important that such calibration is done: if predictive scoring is used to rank events by risk, calibration ought to be such that predicted frequencies track true frequencies \cite{Brier1950Verification,NiculescuMizil2005Predicting,Naeini2015BBQ}.

Targeting matters because small per-event improvements compound. Consider a person resurfacing items over time. If item $i$ is reused with rate $\lambda_i(t)$, the expected number of overexposures over horizon $T$ is
\[
\mathbb{E}[O_i(T)] \;=\; \int_{0}^{T} \lambda_i(t)\, \big(1 - p_c(\mathbf{x}_{i,t})\big)\, p_o(\mathbf{x}_{i,t}) \, dt.
\]
A modest increase in $\tilde{p}_c$ or reduction in $\tilde{p}_o$ on a subset of high-risk events yields meaningful cumulative reductions. The practical question is \emph{which} events to treat under a limited prompt budget. A simple greedy policy sorts candidate events by $c_o\cdot\Delta r(\mathbf{x})$ and selects the top $B$ per day; this is optimal when prompt costs are homogeneous and provides most of the benefit with minimal complexity.

Since experimental access or long-running A/B tests might be infeasible, one could consider an evaluation framework relying solely on the log. First, attach machine-readable provenance wherever possible to the content (for instance, through a lightweight PROV profile \cite{Moreau2013PROVDM}). Second, add three opt-in counters to measure: (1) how often outgoing scope is strictly broader than attached provenance (used as a conservative proxy for overexposure); (2) override rates (i.e., how often posters choose to broaden or narrow); and (3) completion and time-to-post. Third, simulate counterfactual policies by replaying event logs: identify events that would have triggered a badge or sheet under a candidate policy and apply literature-informed priors for $\tilde{p}_c,\tilde{p}_o$ to estimate $\Delta r$. Fourth, report sensitivity bands instead of point estimates to reflect uncertainty, and prefer policies that dominate across a range of plausible priors.

\emph{Table~\ref{tab:cadence_scenarios}} presents an illustrative computation, with completely hypothetical numbers, we present it as an example usage of how those values are being calculated, and how those equations could be used. It shows that selecting only the riskiest bucket of reuse events can minimize estimated overexposures with least friction.

\begin{table}[!t]
\small
\centering
\begin{tabular}{lrrrr}
\toprule
\textbf{Scenario} & \textbf{Events} & $\mathbf{p_c}$ & $\mathbf{p_o}$ & $\mathbb{E}[\text{overexposures}]$ \\
\midrule
Baseline (illustrative) & 360 & 0.57 & 0.57 & $360 \times 0.43 \times 0.57 = 88.24$ \\
Prompt top 30\%         & 108 & \textbf{0.72} & \textbf{0.47} & $108 \times 0.28 \times 0.47 = 14.21$ \\
Unprompted 70\%         & 252 & 0.57 & 0.57 & $252 \times 0.43 \times 0.57 = 61.78$ \\
\midrule
Total with policy       & 360 & 0.62 & 0.54 & $14.21 + 61.78 = \mathbf{76.00}$ \\
\bottomrule
\end{tabular}
\caption{Illustrative cumulative impact under a targeted prompting policy.}
\label{tab:cadence_scenarios}
\end{table}

Instrumenting the patterns seems possible in an ethical manner. Uncertainty proxies (e.g., dwell time over the audience-control after an \emph{Old} cue) can be calculated locally and discarded. Aggregate counters to support evaluation can maintain their counts in differential-privacy bins or coarser time windows. The product requirement is straightforward: to inform people of what the system knows relevant to the decision (the original audience), so that they may accept, broaden, or narrow with very little ceremony. Governance-wise, it is equally simple: document when prompts appear and for what purpose, and then check to ensure pattern libraries do not quietly broaden scope or hide narrow options \cite{Gray2018DarkPatternsSide,Mathur2019DarkPatterns}.

\subsubsection{Operating under uncertainty.}
If directionality is uncertain, we can optimize on \emph{upper bounds} rather than point estimates. Let $p_c, p_o$ lie in intervals $[l_c,u_c],[l_o,u_o]$ derived from literature-informed Beta priors or log-only proxies. Then a safe, cost-sensitive trigger compares the worst-case risk
\[
\mathbb{E}[R]_{\max} = (1 - l_c)\, u_o
\]
against friction $c_f$ via $c_o \cdot \Delta r_{\min} > c_f$, where $\Delta r_{\min}$ uses conservative bounds on intervention lift $(\tilde{p}_c - p_c)$ and $(p_o - \tilde{p}_o)$. This yields a policy that reduces overexposure whenever it is plausibly present, without assuming its magnitude.

\section{Limitations and Future Work}

We use a three-level audience order (\emph{Private}, \emph{Friends}, \emph{Public}) as a scaffold to reason about directionality; many platforms implement richer, non-linear lattices with exceptions (e.g., lists, shared groups, domain-scoped circles). The design patterns generalize—badges can show concrete recipients or examples—but the formal model’s step-distance interpretation becomes more complex. Future work should map drift on realistic audience graphs and study how people reason about partial overlaps or asymmetric expansions (e.g., from “team” to “org-visible”).

Temporal drift likely also interacts with some other cues that we have not foregrounded here: thumbnails, location, comments, and algorithmic surfacing signals (i.e., why is this item resurfacing). Field deployments should investigate whether the presence of clear provenance competes with or complements these cues and if copy such as "Original audience: Friends (May 2024)" changes the social meaning attributed to the re-sharel or its chilling potential. The longer-term social consequences remain an open question: does making prior audiences more salient prompt cultures to reuse with extra care, or does it inadvertently constrain self-presentation \cite{Hogan2010PresentationOfSelf,Bazarova2014SelfDisclosureJCMC,Taddicken2014JCMCPrivacyParadox,Stutzman2013SilentListeners,Kokolakis2017PrivacyParadox,Baruh2017MediatizedPrivacyParadox}?

Beyond the reuse within a single application, this is another extremely important frontier. Lack of interoperable privacy-preserving provenance standards implies that context can easily collapse across apps \cite{sivanSevilla2024contextualIntegrity}. Profiling and reasoning about existing standards such as W3C PROV in terms of the audience claims and use cases would outline safer fallbacks for lossy mappings, thus furnishing designers with workable building blocks \cite{Moreau2013PROVDM}. In some cultures and communities (e.g., stigmatized groups, and adult content creators \cite{bonifacio2025onlyfans}), revealing certain provenance labels is sensitive to whom would be the subject of a research agenda in graduated disclosure for a combined usefulness and safety perspective.

We suggested evaluation involving opt-in instrumentation, log-only simulation, and conservative priors as opposed to personal user studies. This is a feature rather than a limitation once one is dealing with potentially sensitive content and vulnerable populations. Where practical and when ethical review allows, field trials can in fact calibrate priors and test portability with strong privacy guarantees. When experimentation is not viable, we suggest presenting sensitivity bands and focus on interventions that dominate over ranges of $p_c,p_o$ consistent with the literature \cite{Acquisti2015PrivacyScience,Brandimarte2013MisplacedConfidences,Tversky1974Judgment,Kahneman2011Thinking}.

Finally, beyond social media, analogous drift appears in document management, messaging, and media vaults (e.g., private galleries \cite{Geeng2024Vault}). Adapting badges and restoration prompts to those domains, with domain-specific audience ontologies and higher-stakes defaults, is promising. Richer UI exemplars (e.g., mockups vetted with accessibility experts) and open-source pattern libraries could accelerate adoption.

\section{Conclusion}

This paper promotes a view of privacy decision-making that recognizes time as a factor. As time passes and content is shared, people reference gist instead of verbatim memories of what they previously shared, producing more errors towards wider audiences. We draw on theory from memory, contextual integrity, and usable privacy to coin this concept \emph{temporal drift of privacy recall} and operationalize it into tangible, low-burden design interventions. In particular, we put forward three complementary patterns—portable provenance badges, context-restoration prompts as needed, and narrow-first initialization when provenance is absent—that convert recall to recognition in the exact moment that materials are reused and retain fluidity and user agency.

To inform deployment decisions, we present a cost-sensitive risk model that defines the potential for overexposure as a product of recall error and direction of error, and we provide a log-only evaluation strategy that leverages privacy-preserving instrumentation, counterfactual replays, and sensitivity bands rather than conventional studies with users. Altogether these provide a feasible alternative for teams wanting to deploy improvements, while working within the constraints of scale, ethics, and limited experimental access. We also surface governance and accessibility challenges: make provenance legible and avoid mining sensitive groups by default; compute uncertainty in a local manner and discard raw signals; avoid dark patterns; and internationalize copy and interaction.

Our argument can be extended beyond social media. The same type of drift can occur in shared drives, LMSs, Patient Portals, and media vaults when it is necessary to recall prior recipients at the time of reusing. Light touch machine-readable provenance, revealed mapping constraints across platforms, and falling back conservatively are actionable steps toward safer portability.

\bibliographystyle{ACM-Reference-Format}
\bibliography{refs}

\appendix

\end{document}